\documentclass[english,conference]{IEEEtran}
\PassOptionsToPackage{ruled, linesnumbered}{algorithm2e}
\usepackage[T1]{fontenc}
\usepackage[latin9]{inputenc}
\usepackage{color}
\usepackage{mathrsfs}
\usepackage{algorithm2e}
\usepackage{amsmath}
\usepackage{amssymb}
\usepackage{graphicx}

\makeatletter
\usepackage{cite, balance,epstopdf}

\usepackage{amsthm}

\theoremstyle{remark}

\usepackage{algorithm2e}

\DeclareMathOperator{\tr}{tr}

\DeclareMathOperator{\diag}{diag}
\DeclareMathOperator{\maximize}{maximize}
\DeclareMathOperator{\st}{subject~to}
\DeclareMathOperator{\vecd}{\mathrm{vec_d}}
\newcommand{\herm}{^{{\dagger}}}
\newcommand{\trans}{^{\mathsf{T}}}
\DeclareMathOperator{\argmin}{argmin}

\allowdisplaybreaks
\renewcommand{\fnum@figure}{Fig. \thefigure}
\makeatother

\usepackage{babel}
\begin{document}
\title{\huge{On the Energy-Efficiency Maximization for IRS-Assisted MIMOME
Wiretap Channels}}
\author{\IEEEauthorblockN{Anshu~Mukherjee\IEEEauthorrefmark{1}, Vaibhav~Kumar\IEEEauthorrefmark{1},
Derrick~Wing~Kwan~Ng\IEEEauthorrefmark{2}, and Le-Nam~Tran\IEEEauthorrefmark{1}}\IEEEauthorblockA{\IEEEauthorrefmark{1}School of Electrical and Electronic Engineering,
University College Dublin, Belfield, Dublin 4, Ireland\\
\IEEEauthorrefmark{2}School of Electrical Engineering and Telecommunications,
University of New South Wales, NSW 2052, Australia\\
Email: anshu.mukherjee@ucdconnect.ie, vaibhav.kumar@ieee.org, w.k.ng@unsw.edu.au,
nam.tran@ucd.ie}}

\maketitle
{\let\thefootnote\relax\footnotetext{This publication has emanated from research conducted with the financial support of Science Foundation Ireland (SFI) and is co-funded under the European Regional Development Fund under Grant Number 17/CDA/4786.}}
\begin{abstract}
Security and energy efficiency have become crucial features in the
modern-era wireless communication. In this paper, we consider an energy-efficient
design for intelligent reflecting surface (IRS)-assisted multiple-input
multiple-output multiple-eavesdropper (MIMOME) wiretap channels (WTC).
Our objective is to jointly optimize the transmit covariance matrix
and the IRS phase-shifts to maximize the secrecy energy efficiency
(SEE) of the considered system subject to a secrecy rate constraint
at the legitimate receiver. To tackle this challenging non-convex
problem in which the design variables are coupled in the objective
and the constraint, we propose a \textit{penalty dual decomposition
based alternating gradient projection (PDDAPG)} method to obtain an
efficient solution. We also show that the computational complexity
of the proposed algorithm grows only \textit{linearly} with the number
of reflecting elements at the IRS, as well as with the number of antennas
at transmitter/receivers' nodes. Our results confirm that using an
IRS is helpful to improve the SEE of MIMOME WTC compared to its no-IRS
counterpart only when the power consumption at IRS is small. In particular,
and a large-sized IRS is not always beneficial for the SEE of a MIMOME
WTC.
\end{abstract}


\section{Introduction}

As the fifth-generation (5G) wireless communication networks are being
rolled out by different mobile service providers worldwide, the research
community has started looking for the next breakthrough in beyond-5G
(B5G) cellular standard requiring new goals of B5G. The intelligent
reflecting surface (IRS)~\cite{RIS6gMag} is one such a technology
that has the potential to cater to the demand of supporting an exponentially-increasing
number of devices within the extremely-congested sub-6~GHz spectrum.
It has been verified that IRSs can significantly enhance the spectral
and/or energy efficiency of a wireless communication system~\cite{Stefan2021MIMORIS,EEIRS2019Debbah}.
Meanwhile, the unprecedentedly increased dependence of our day-to-day
life on wireless communication over the last decades has raised serious
concerns about the security- and privacy-related issues of these services~\cite{Derrick2017MIMOMEWTCSurvey}.
On this front, different from the technologies that are implemented
on the higher layers, the IRSs have shown considerable potential to
facilitate secure communication from the physical-layer perspective~\cite{Anshu_Milcom,Vaibhav_Globecom}.

Although achievable (secrecy) rate has been an important figure of
merit in previous-generation wireless standards, maximizing the energy
efficiency (EE) has become another crucial performance measure for
the next-generation wireless networks~\cite{JieXu2013MIMOEnrgEff,BhargavaMag}.
Some of recent publications related to the EE maximization in IRS-assisted
wireless systems include~\cite{EEIRS2019Debbah,EE_D2D_YCLiang,EEmaximization_AtaKhalili,EE_DistributedRIS}.
On the other hand, there are only a couple of works available dealing
with the problem of EE maximization for IRS-assisted \textit{secure}
communications. For instance, the authors in~\cite{Secure_EE_CognitiveRadio}
considered the problem of secrecy energy efficiency (SEE) maximization
for an IRS-assisted multi-input single-output single-eavesdropper
(MISOSE) spectrum sharing system, where a suboptimal solution was
obtained adopting an alternating optimization (AO) based approach,
in conjunction with an iterative penalty-function-based algorithm
and a difference-of-convex (DC) functions method. For a non-cognitive
MISOSE system along with a dedicated friendly jammer, the authors
in~\cite{Secure_EE_Jamming} considered the problem of SEE maximization,
where a suboptimal solution was obtained using semidefinite programming
(SDP) and Dinkelbach's method.

Although MISO systems are of particular interest in many applications,
including those in the Internet-of-Things (IoT), MIMO systems remain
to be an integral part of B5G standard. It is important to note that
since MIMO systems enjoy a larger diversity and multiplexing gain
over that of MISO systems, they are more suitable for applications
that require very-high data rate, ultra-reliability and secrecy as
well, such as healthcare and military applications. The existing solution(s)
for the SEE maximization mentioned above for IRS-assisted MISOSE WTC
systems are not directly applicable to a general MIMOME setting. Therefore,
in this paper we consider the problem of SEE maximization in an IRS-assisted
MIMOME system, which to the best of our knowledge has not been investigated
earlier. The formulated optimization problem is highly non-convex,
making it challenging to solve, and therefore deserves a separate
thorough study. The main contributions in this paper are listed below:
\begin{figure}[t]
\begin{centering}
\includegraphics[width=0.6\columnwidth]{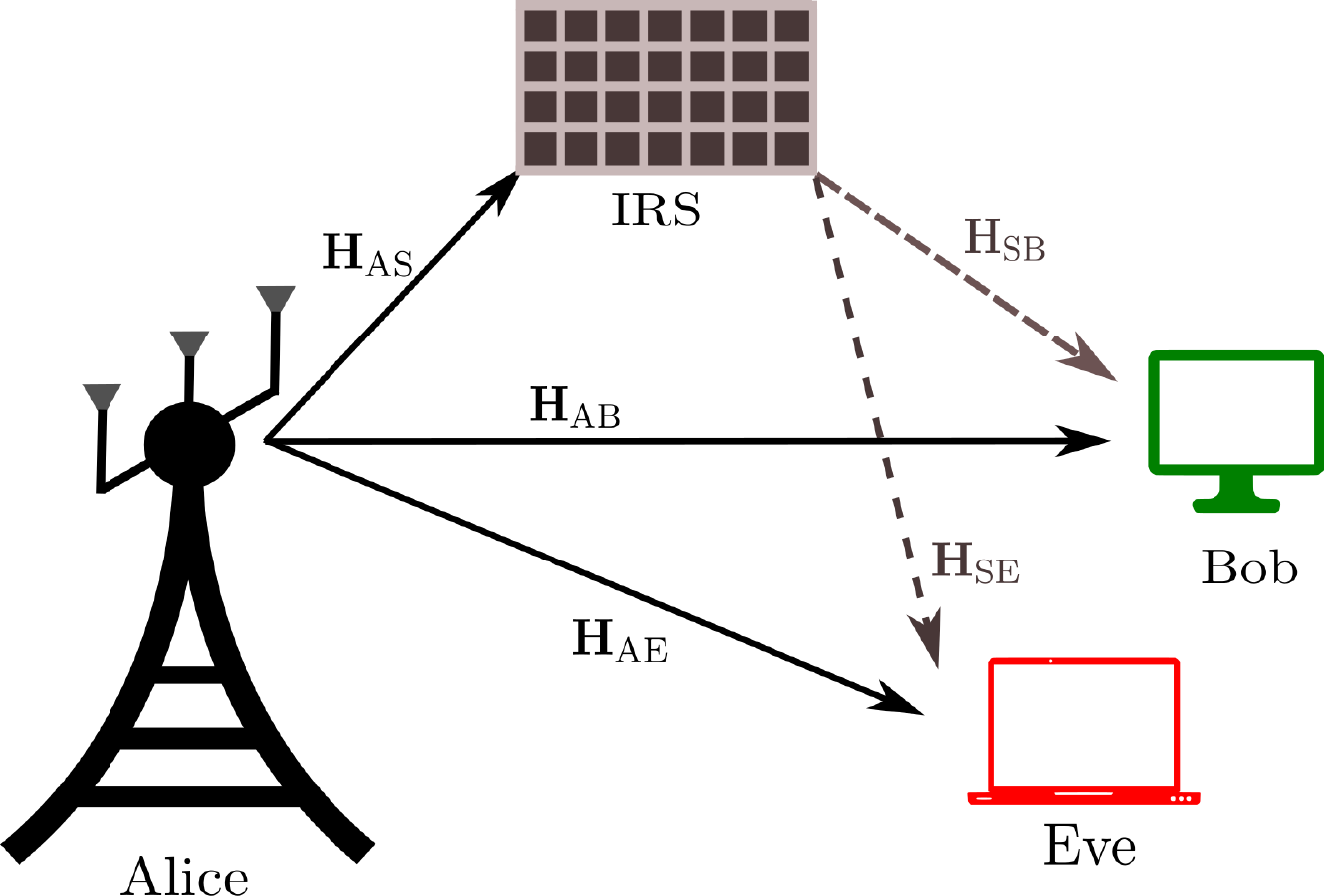}
\par\end{centering}
\caption{An IRS-MIMOME system.}
\label{fig:SysMod}
\end{figure}

\begin{itemize}
\item We propose a simple, yet efficient, numerical solution for the active
and passive beamforming design problem to maximize the SEE of the
IRS-MIMOME system. In particular, we propose a a penalty dual decomposition
based alternating gradient projection (PDDAGP) method to obtain a
stationary solution to the formulated non-convex optimization problem.
\item We also provide a detailed computational complexity analysis\textcolor{red}{{}
}for the proposed PDDAPG method which confirms that the complexity
of our proposed method grows only linearly with respect to (w.r.t.)
the number of reflecting elements at the IRS, as well as the number
of antennas at the transmitter/receivers.
\item We provide extensive numerical results to evaluate the performance
of the proposed method. Additionally, we compare the performance of
the proposed method with a baseline scheme adopting zero-forcing method
for the input covariance and a Gaussian randomization for the IRS
phase shifts. Especially, for the special case of the IRS-MISOSE system,
our proposed algorithm is shown to be superior to that proposed in~\cite{Secure_EE_Jamming},
both in terms of average SEE and average runtime.
\end{itemize}
\textit{Notations:} We use bold lowercase and uppercase letters to
denote column vectors and matrices, respectively. The Hermitian transpose
and (ordinary) transpose operators are respectively denoted by $(\cdot)\herm$
and $(\cdot)\trans$. We use $\mathbb{C}^{M\times N}$ $(\mathbb{R}^{M\times N})$
to denote the vector space of complex-valued (real-valued) matrices
of size $M\times N$. $\diag(\mathbf{\cdot})$ denotes the square
diagonal matrix and $\vecd(\mathbf{Y})$ represents the column vector
with elements taken from the main diagonal of $\mathbf{Y}$. $\mathbf{I}$
specify an identity matrix. We denote the trace, determinant, and
Frobenius norm of the matrix $\mathbf{Y}$ by $\tr(\boldsymbol{\mathbf{Y}})$,
$\left|\mathbf{Y}\right|$, and $\Vert\mathbf{Y}\Vert$, respectively.
The complex-valued gradient of a function $f(\cdot)$ with respect
to (w.r.t.) $\mathbf{X}^{*}$ is denoted by $\frac{\partial}{\partial\mathbf{X}^{\ast}}f(\cdot)=\frac{1}{2}\Bigl(\frac{\partial f(\cdot)}{\partial\Re(\mathbf{X})}+j\frac{\partial f(\cdot)}{\partial\Im(\mathbf{X})}\Bigr)$,
and $\odot$ represents the Hadamard (i.e. entry-wise) product. By
$\mathbf{A}\succeq(\textrm{resp.}\ \succ)\ \mathbf{B}$ we mean $\mathbf{A}-\mathbf{B}$
is positive semidefinite (resp. definite). We define $[x]_{+}\triangleq\max\{x,0\}$.
The Euclidean projection of $\mathbf{y}$ onto a feasible set $\mathcal{Y}$
is denoted by $\Pi_{\mathcal{Y}}(\mathbf{y})\triangleq\argmin_{\tilde{\mathbf{y}}\in\mathcal{Y}}\left\Vert \tilde{\mathbf{y}}-\mathbf{y}\right\Vert $.
$\mathcal{O}(\cdot)$ represents the Bachmann--Landau notation.

\section{System Model and Problem Formulation}

In this section, we describe the system model and formulate the SEE
maximization problem for the system under consideration.

\subsection{System Model}

Consider an IRS-MIMOME system shown in~Fig.~\ref{fig:SysMod}, consisting
of one transmitter (Alice), one legitimate receiver (Bob), and one
eavesdropper (Eve). The number of antennas at Alice, Bob, and Eve
are denoted by $N_{\mathrm{A}}$, $N_{\mathrm{B}}$, and $N_{\mathrm{E}}$,
respectively, and the IRS is assumed to be manufactured with $N_{\mathrm{S}}$
low-cost passive reflecting elements. The complex-valued channel matrices
for the Alice-IRS, IRS-Bob, IRS-Eve, Alice-Bob and Alice-Eve links
are denoted by $\mathbf{H}_{\mathrm{AS}}\in\mathbb{C}^{N_{\mathrm{S}}\times N_{\mathrm{A}}}$,
$\mathbf{H}_{\mathrm{SB}}\in\mathbb{C}^{N_{\mathrm{B}}\times N_{\mathrm{S}}}$,
$\mathbf{H}_{\mathrm{SE}}\in\mathbb{C}^{N_{\mathrm{E}}\times N_{\mathrm{S}}}$,
$\mathbf{H}_{\mathrm{AB}}\in\mathbb{C}^{N_{\mathrm{B}}\times N_{\mathrm{A}}}$,
and $\mathbf{H}_{\mathrm{AE}}\in\mathbb{C}^{N_{\mathrm{E}}\times N_{\mathrm{A}}}$,
respectively. Following the arguments in~\cite{Stefan2021MIMORIS,EEIRS2019Debbah,Anshu_Milcom,Vaibhav_Globecom,EE_D2D_YCLiang,EEmaximization_AtaKhalili,EE_DistributedRIS,Secure_EE_CognitiveRadio,Vaibhav_SpectrumSharing},
it is assumed that all these channel matrices are quasi-static and
perfectly known at all of the nodes.\footnote{The results in this paper serve as theoretical performance upper bounds
for the IRS-MIMOME system with imperfect channel state information in practice.}

The received signals at Bob and Eve are, respectively, expressed as
\begin{equation}
\begin{aligned}\mathbf{y}_{\mathrm{B}}= & (\mathbf{H}_{\mathrm{AB}}+\mathbf{H}_{\mathrm{SB}}\mathbf{\mathbf{Z}(\boldsymbol{\theta})}\mathbf{H}_{\mathrm{AS}})\mathbf{x}+\mathbf{w}_{\mathrm{B}},\\
\mathbf{y}_{\mathrm{\mathrm{E}}}= & (\mathbf{H}_{\mathrm{AE}}+\mathbf{H}_{\mathrm{SE}}\mathbf{Z}(\boldsymbol{\theta})\mathbf{H}_{\mathrm{AS}})\mathbf{x}+\mathbf{w}_{\mathrm{E}},
\end{aligned}
\label{eq:RxSignals}
\end{equation}
where $\mathbf{x}\in\mathbb{C}^{N_{\mathrm{A}}\times1}$ is the transmitted
signal vector from Alice (intended for Bob); $\mathbf{w}_{\mathrm{B}}\sim\mathcal{CN}(\boldsymbol{0},\sigma_{\mathrm{B}}^{2}\mathbf{I})$
and $\mathbf{w}_{\mathrm{E}}\sim\mathcal{CN}(\boldsymbol{0},\sigma_{\mathrm{E}}^{2}\mathbf{I})$
are the additive white Gaussian noise (AWGN) at Bob and Eve, respectively.
For ease of presentation, in the rest of the paper we assume $\sigma_{\mathrm{B}}=\sigma_{\mathrm{E}}=\sigma=\sqrt{\mathscr{N}_{0}B}$,
where $\mathscr{N}_{0}$ is noise power spectral density and $B$
is the signal bandwidth. With a slight abuse of notation and without
loss of generality, in the sequel of the paper, we normalize the involving
channels appropriately with the noise power, i.e., we define $\mathbf{\mathbf{H}_{\mathrm{AB}}}\leftarrow\tfrac{1}{\sigma}\mathbf{\mathbf{H}_{\mathrm{AB}}}$,
$\mathbf{\mathbf{H}_{\mathrm{SB}}}\leftarrow\tfrac{1}{\sigma}\mathbf{\mathbf{H}_{\mathrm{SB}}}$,
$\mathbf{\mathbf{H}_{\mathrm{AE}}}\leftarrow\tfrac{1}{\sigma}\mathbf{\mathbf{H}_{\mathrm{AE}}}$,
and $\mathbf{\mathbf{H}_{\mathrm{SE}}}\leftarrow\tfrac{1}{\sigma}\mathbf{\mathbf{H}_{\mathrm{SE}}}$,
and thus the resulting equivalent noise has a unit variance. In~(\ref{eq:RxSignals}),
$\mathbf{Z}(\boldsymbol{\theta})\triangleq\diag(\boldsymbol{\theta})$,
where $\boldsymbol{\theta}\triangleq[\theta_{1},\theta_{2},\dots,\theta_{N}]\trans\in\mathbb{C}^{N_{\mathrm{S}}\times1}$,
$\theta_{i}=e^{j\phi_{i}}$, $i\in\mathscr{N}_{\mathrm{S}}\triangleq\left\{ 1,2,\ldots,N_{\mathrm{S}}\right\} $,
and $\phi_{i}\in[0,2\pi)$ denotes the phase shift induced by the
$i$-th reflecting element at the IRS.\footnote{Although various models for IRS phase-shifts have been introduced
recently~\cite{OptMag}, the unit-modulus model is the most frequently
used in literature~\cite{Stefan2021MIMORIS,EEIRS2019Debbah,Anshu_Milcom,Vaibhav_Globecom,EE_D2D_YCLiang,EEmaximization_AtaKhalili,EE_DistributedRIS,Secure_EE_CognitiveRadio,Vaibhav_SpectrumSharing} for research investigation.}

\subsection{Problem Formulation}

Denoting the transmit covariance matrix at Alice by $\mathbf{X}\triangleq\mathbb{E}\left\{ \mathbf{x}\mathbf{x}\herm\right\} \succeq\boldsymbol{0}$,
the achievable secrecy rate (in nats/s/Hz) between Alice and Bob is
given by
\begin{equation}
C(\mathbf{X},\boldsymbol{\theta})=[\ln\bigl|\mathbf{I}+\mathbf{H}_{\mathrm{B}}\mathbf{X}\mathbf{H}_{\mathrm{B}}\herm\bigr|-\ln\bigl|\mathbf{I}+\mathbf{H}_{\mathrm{E}}\mathbf{X}\mathbf{H}_{\mathrm{E}}\herm\bigr|]_{+},\label{eq:SecrecyRateDef}
\end{equation}
where $\mathbf{H}_{\mathrm{B}}\triangleq(\mathbf{H}_{\mathrm{AB}}+\mathbf{H}_{\mathrm{SB}}\mathbf{\mathbf{Z}(\boldsymbol{\theta})}\mathbf{H}_{\mathrm{AS}})$,
$\mathbf{H}_{\mathrm{E}}\triangleq(\mathbf{H}_{\mathrm{AE}}+\mathbf{H}_{\mathrm{SE}}\mathbf{Z}(\boldsymbol{\theta})\mathbf{H}_{\mathrm{AS}})$.
On the other hand, the total power consumption to achieve the secrecy
rate given in~(\ref{eq:SecrecyRateDef}) is expressed as~(c.f.~\cite{EEIRS2019Debbah})
\begin{equation}
P_{\mathrm{total}}=\frac{\tr(\mathbf{X})}{\alpha}+P_{\mathrm{A}}+P_{\mathrm{S}}+P_{\mathrm{B}},\label{eq:PowerConsumptionModel}
\end{equation}
where $P_{\mathrm{A}}$ , $P_{\mathrm{S}}$, and $P_{\mathrm{B}}$
are the constant circuit power consumption at Alice, IRS and Bob,
respectively, and $\alpha\in(0,1]$ is the power amplifier efficiency
at Alice. Following the arguments in~\cite{EEIRS2019Debbah}, the
power consumption at the IRS is modeled as $P_{\mathrm{S}}=N_{\mathrm{S}}P_{\mathrm{e}}$,
where $P_{\mathrm{e}}$ is the circuit power consumption at each of
the reflecting element in the IRS. Note that in~(\ref{eq:PowerConsumptionModel}),
we do not consider the circuit power consumption at Eve. Such a scenario
is justified where Eve is a node external to the main system. Then
the SEE (in nats/s/Joule) for the IRS-MIMOME system can be defined
as 
\begin{equation}
\mathcal{E}(\mathbf{X},\boldsymbol{\theta})\triangleq B\mathsf{E}(\mathbf{X},\boldsymbol{\theta})=B\frac{C(\mathbf{X},\boldsymbol{\theta})}{P_{\mathrm{total}}},\label{eq:SEEDef}
\end{equation}
where $B$ is the total available bandwidth and $\mathsf{E}(\mathbf{X},\boldsymbol{\theta})$
is the SEE in nats/s/Hz/Joule. In this paper, our objective is to
maximize the SEE of the IRS-MIMOME system, the corresponding optimization
problem can be formulated as follows: 
\begin{subequations}
\label{eq:mainOptProb}
\begin{align}
\underset{\mathbf{X},\,\boldsymbol{\theta}}{\mathrm{maximize}} & \ \Big\{\mathsf{E}(\mathbf{X},\boldsymbol{\theta})=\frac{C(\mathbf{X},\boldsymbol{\theta})}{P_{\mathrm{total}}}\Big\}\label{eq:MainProblem}\\
\st & \ C(\mathbf{X},\boldsymbol{\theta})\geq C_{\mathrm{th}},\label{eq:SecThresConstraint}\\
 & \ \tr(\mathbf{X})\leq P_{\max},\label{eq:TPC}\\
 & \ |\theta_{i}|=1,\forall i\in\mathscr{N}_{\mathrm{S}},\label{eq:UMC}
\end{align}
\end{subequations}
 where $P_{\max}$ is the maximum transmit power budget at Alice,
and $C_{\mathrm{th}}$ is the threshold secrecy rate to maintain a
minimum required quality of service (QoS) at Bob. Note that $\mathsf{E}(\mathbf{X},\boldsymbol{\theta})$
is the SEE of the system under consideration, the constraint in~(\ref{eq:SecThresConstraint})
ensures the secrecy QoS at Bob, and those in~(\ref{eq:TPC}) and~(\ref{eq:UMC})
refer to the transmit power constraint at Alice and unit-modulus constraints
at the IRS, respectively. We further define the feasible set for the
optimization variables $\mathbf{X}$ and $\boldsymbol{\theta}$ as
$\mathcal{S}_{\mathbf{X}}\triangleq\{\mathbf{X}\in\mathbb{C}^{N_{\mathrm{A}}\times N_{\mathrm{A}}}:\mathbf{X}\succeq\mathbf{0},\tr(\mathbf{X})\leq P_{\max}\}$
and $\mathcal{S}_{\boldsymbol{\theta}}\triangleq\{\boldsymbol{\theta}\in\mathbb{C}^{N_{\mathrm{S}}\times1}:|\theta_{i}|=1,i\in\mathscr{N}_{\mathrm{S}}\}$,
respectively.

\section{Proposed Solution}

\subsection{Algorithm Description}

It can be observed that the problem in~(\ref{eq:mainOptProb}) is
non-convex due to the coupling of $\mathbf{X}$ and $\boldsymbol{\theta}$
in both~(\ref{eq:MainProblem}) and~(\ref{eq:SecThresConstraint}),
and the non-convexity of the constraints in~(\ref{eq:UMC}). In order
to solve a similar optimization problem for a system consisting of
a multi-antenna Alice, one single-antenna Bob, multiple single-antenna
Eve, and one multi-antenna friendly jammer, the authors in~\cite{Secure_EE_Jamming}
proposed an AO-based algorithm using SDP and Dinkelbach's method.
It is important to note that due to the existing coupling of the optimization
variables in both objective function and optimization constraints,
the proposed AO-based algorithm in~\cite{Secure_EE_Jamming} cannot
guarantee a stationary solution to the optimization problem involving
the use of Gaussian randomization to recover the rank-one matrix,
and thus normally results in an inferior system performance. Furthermore,
the complexity of the proposed solution in~\cite{Secure_EE_Jamming}
is excessively high due to the use of SDP. Motivated by these drawbacks,
in this paper, we propose a low-complexity PDDAPG method to find a
stationary solution to~(\ref{eq:mainOptProb}). A similar method
was recently shown to be effective for the achievable rate maximization
problem in an IRS-assisted MIMO underlay spectrum sharing system in~\cite{Vaibhav_SpectrumSharing}.

In order to deal with the non-convex coupling constraint in~(\ref{eq:SecThresConstraint}),
we adopt the penalty dual decomposition method~\cite{PDDTheory2020Shi}.
For this purpose, we define $f(\mathbf{X},\boldsymbol{\theta},C_{\mathrm{th}},\varsigma)\triangleq C_{\mathrm{th}}-C(\mathbf{X},\boldsymbol{\theta})+\varsigma$.
It is straightforward to note that for some $\varsigma\geq0$,~(\ref{eq:SecThresConstraint})
is equivalent to $f(\mathbf{X},\boldsymbol{\theta},C_{\mathrm{th}},\varsigma)=0$.
Following the arguments in~\cite{PDDTheory2020Shi}, for a given
Lagrangian multiplier $\nu$ and a penalty parameter $\omega\geq0$,
an augmented Lagrangian function corresponding to~(\ref{eq:MainProblem})
can be defined as follows:
\begin{align}
\hat{\mathsf{E}}_{\nu,\omega}(\mathbf{X},\boldsymbol{\theta},\varsigma) & \triangleq\mathsf{E}(\mathbf{X},\boldsymbol{\theta})-\nu f(\mathbf{X},\boldsymbol{\theta},C_{\mathrm{th}},\varsigma)\nonumber \\
 & \quad\quad-\frac{\omega}{2}f^{2}(\mathbf{X},\boldsymbol{\theta},C_{\mathrm{th}},\varsigma).\label{eq:AugObj}
\end{align}
Therefore, for fixed $\nu$ and $\omega$, we arrive at the following
equivalent optimization problem:

\begin{subequations}
\label{eq:OptProb-1}
\begin{align}
\underset{\mathbf{X},\boldsymbol{\theta},\varsigma}{\maximize}\  & \hat{\mathsf{E}}_{\nu,\omega}(\mathbf{X},\boldsymbol{\theta},\varsigma)\label{eq:OptObj-1}\\
\st\  & \varsigma\geq0,\eqref{eq:TPC},\eqref{eq:UMC}.\label{eq:varSigmaConstraint}
\end{align}
\end{subequations}
It is noteworthy that the coupling of $\mathbf{X}$ and $\boldsymbol{\theta}$
is now included in the augmented objective and the constraints are
decoupled in~(\ref{eq:OptProb-1}). Therefore, to obtain a stationary
solution to~(\ref{eq:OptProb-1}), we apply a simple, yet efficient,
numerical technique based on alternating gradient projection (APG)
method.\footnote{Following the arguments in~\cite{PDDTheory2020Shi}, it can be shown
that a stationary solution to~(\ref{eq:OptProb-1}) is indeed a
stationary solution to the original problem in~(\ref{eq:mainOptProb})
at the convergence.} For this purpose, we first find $\nabla_{\mathbf{X}}\hat{\mathsf{E}}_{\nu,\omega}(\mathbf{X},\boldsymbol{\theta},\varsigma)$
as follows:
\begin{align}
 & \nabla_{\mathbf{X}}\hat{\mathsf{E}}_{\nu,\omega}(\mathbf{X},\boldsymbol{\theta},\varsigma)=\big[\tfrac{1}{P_{\mathrm{total}}}+\nu+\omega f(\mathbf{X},\boldsymbol{\theta},C_{\mathrm{th}},\varsigma)\big]\nonumber \\
 & \qquad\qquad\qquad\qquad\times\nabla_{\mathbf{X}}C(\mathbf{X},\boldsymbol{\theta})-\tfrac{C(\mathbf{X},\boldsymbol{\theta})}{P_{\mathrm{total}}^{2}}\nabla_{\mathbf{X}}\tr(\mathbf{X}).\label{eq:gradXdef}
\end{align}
Using~(\ref{eq:gradXdef}) and ~\cite[eqns. (6.207), (6.195) and Table 4.3]{DerivativeBook},
a closed-form expression for $\nabla_{\mathbf{X}}\hat{\mathsf{E}}_{\nu,\omega}(\mathbf{X},\boldsymbol{\theta},\varsigma)$
is given by~(\ref{eq:gradXclosed}), shown at the top of the next
page. On the other hand, using~\cite[eqn. (17a)]{Stefan2021MIMORIS}
and~\cite[Table 4.3 and eqn. (6.153)]{DerivativeBook}, a closed-form
expression for $\nabla_{\mathbf{\boldsymbol{\theta}}}\hat{\mathsf{E}}_{\nu,\omega}(\mathbf{X},\boldsymbol{\theta},\varsigma)$
is given by~(\ref{eq:gradThetaClosed}), shown at the top of the
next page.
\begin{figure*}[t]
\begin{equation}
\nabla_{\mathbf{X}}\hat{\mathsf{E}}_{\nu,\omega}(\mathbf{X},\boldsymbol{\theta},\varsigma)=\big[\tfrac{1}{P_{\mathrm{total}}}+\nu+\omega f(\mathbf{X},\boldsymbol{\theta},C_{\mathrm{th}},\varsigma)\big]\big[\mathbf{H}_{\mathrm{B}}\herm\big(\mathbf{I}+\mathbf{H}_{\mathrm{B}}\mathbf{X}\mathbf{H}_{\mathrm{B}}\herm\big)^{-1}\mathbf{H}_{\mathrm{B}}-\mathbf{H}_{\mathrm{E}}\herm\big(\mathbf{I}+\mathbf{H}_{\mathrm{E}}\mathbf{X}\mathbf{H}_{\mathrm{E}}\herm\big)^{-1}\mathbf{H}_{\mathrm{E}}\big]-\tfrac{C(\mathbf{X},\boldsymbol{\theta})}{P_{\mathrm{total}}^{2}}\mathbf{I}.\label{eq:gradXclosed}
\end{equation}
\begin{multline}
\nabla_{\boldsymbol{\theta}}\hat{\mathsf{E}}_{\nu,\omega}(\mathbf{X},\boldsymbol{\theta},\varsigma)=\big[\tfrac{1}{P_{\mathrm{total}}}+\nu+\omega f(\mathbf{X},\boldsymbol{\theta},C_{\mathrm{th}},\varsigma)\big]\times\vecd\big\{\mathbf{H}_{\mathrm{SB}}\herm\big(\mathbf{I}+\mathbf{H}_{\mathrm{B}}\mathbf{X}\mathbf{H}_{\mathrm{B}}\herm\big)^{-1}\mathbf{H}_{\mathrm{B}}\mathbf{X}\mathbf{H}_{\mathrm{AS}}\herm\\
-\mathbf{H}_{\mathrm{SE}}\herm\big(\mathbf{I}+\mathbf{H}_{\mathrm{E}}\mathbf{X}\mathbf{H}_{\mathrm{E}}\herm\big)^{-1}\mathbf{H}_{\mathrm{E}}\mathbf{X}\mathbf{H}_{\mathrm{AS}}\herm\big\}.\label{eq:gradThetaClosed}
\end{multline}
\hrulefill
\end{figure*}

\begin{algorithm}[h]
\caption{Gradient Projection Algorithm to solve~(\ref{eq:OptProb-1}) for
fixed $\nu$ and $\omega$.}

\label{algoGP}

\KwIn{ $\mathbf{X}_{0}$, $\boldsymbol{\theta}_{0}$, $\varsigma_{0}$,
$\tau_{0}$, $\mu_{0}$, $\nu$, $\omega$ }

\KwOut{ $\mathbf{X}_{n}$, $\boldsymbol{\theta}_{n}$}

$n\leftarrow1$

\Repeat{$\mathrm{convergence}$ }{

$\!\!\mathbf{X}_{n}=\Pi_{\mathcal{S}_{\mathbf{X}}}(\hat{\mathbf{X}}_{n}\triangleq\mathbf{X}_{n\!-\!1}+\!\tau_{n}\nabla_{\mathbf{X}}\hat{\mathsf{E}}_{\nu,\omega}(\mathbf{X}_{n-1},\boldsymbol{\theta}_{n-1},\varsigma_{n-1}))$\;

$\!\!\boldsymbol{\theta}_{n}=\Pi_{\mathcal{S}_{\boldsymbol{\theta}}}(\hat{\boldsymbol{\theta}}_{n}\triangleq\boldsymbol{\theta}_{n\!-\!1}\!+\!\mu_{n}\nabla_{\boldsymbol{\theta}}\hat{\mathsf{E}}_{\nu,\omega}(\mathbf{X}_{n},\boldsymbol{\theta}_{n-1},\varsigma_{n-1}))$\;

$\!\!\varsigma_{n}=\max\{0,C(\mathbf{X}_{n},\boldsymbol{\theta}_{n})-C_{\mathrm{th}}\}$\;

$\!\!$$n\leftarrow n+1$\;

}
\end{algorithm}

The gradient projection algorithm to solve~(\ref{eq:OptProb-1})
for fixed $\nu$ and $\omega$ is given in~\textbf{Algorithm~\ref{algoGP}},
where $\tau_{n}$ and $\mu_{n}$ are the step size corresponding to
$\mathbf{X}$ and $\boldsymbol{\theta}$, respectively. Moreover,
in line~3, for a given $\hat{\mathbf{X}}_{n}$, its projection onto
the set $\mathcal{S}_{\mathbf{X}}$, i.e., $\Pi_{\mathcal{S}_{\mathbf{X}}}(\hat{\mathbf{X}}_{n})$
can be shown to admit a water-filling solution. On the other hand,
in line~4, for a given $\hat{\boldsymbol{\theta}}_{n}=[\hat{\theta}_{n,1},\hat{\theta}_{n,2},\ldots,\hat{\theta}_{n,N_{\mathrm{S}}}]\trans$,
its projection onto the set $\mathcal{S}_{\boldsymbol{\theta}}$,
i.e., $\Pi_{\mathcal{S}_{\boldsymbol{\theta}}}(\hat{\boldsymbol{\theta}}_{n})$
is given by $[\theta_{n,1},\theta_{n,2},\ldots,\theta_{n,N_{\mathrm{S}}}]\trans$,
where 
\begin{equation}
\theta_{n,i}\!=\!\left\{ \!\!\begin{array}{cc}
\hat{\theta}_{n,i}/|\hat{\theta}_{n,i}|, & \mathrm{if}\ \hat{\theta}_{n,i}\neq0\\
\exp(j\phi),\phi\in[0,2\pi), & \mathrm{otherwise}
\end{array}\right.,\forall i\in\mathscr{N}_{\mathrm{S}}.\label{eq:ProjTheta}
\end{equation}
Note that~(\ref{eq:ProjTheta}) ensures $|\theta_{n,i}|=1,\forall i\in\mathscr{N}_{\mathrm{S}}$,
and when $|\hat{\theta}_{n,i}|=0$, $\theta_{n,i}$ is chose randomly
from the interval $[0,2\pi)$. After updating $\mathbf{X}$ and $\boldsymbol{\theta}$
in~\textbf{Algorithm~\ref{algoGP}}, we update $\varsigma$ in
line~6. Appropriate values of $\tau_{n}$ and $\mu_{n}$ in each
iteration can be obtained using a \textit{backtracking line search}
routine as suggested in~\cite[Sec. IV-C]{Stefan2021MIMORIS}. Once
the convergence is achieved in~\textbf{Algorithm~\ref{algoGP}},
we update the Lagrange multiplier $\nu$ and the penalty parameter
$\omega$. The overall description of the proposed PDDAGP method to
find a stationary solution to~(\ref{eq:OptProb-1}) is outlined in~\textbf{Algorithm~\ref{algoPDDAGP}}.
The convergence of~\textbf{Algorithm~\ref{algoPDDAGP}} can be
proved following the similar line of arguments as provided in~\cite[Sec. III-C]{Vaibhav_SpectrumSharing}.

\begin{algorithm}[t]
\caption{The PDDAGP Method.}

\label{algoPDDAGP}

\KwIn{$\mathbf{X}_{0}$, $\boldsymbol{\theta}_{0}$, $\varsigma_{0}$,
$\tau_{0}$, $\mu_{0}$, $\nu$, $\omega$, $\eta<1$ }

\KwOut{ $\mathbf{X}^{\star},\boldsymbol{\theta}^{\star}$}

\Repeat{$\mathrm{convergence}$}{

Solve problem~(\ref{eq:OptProb-1}) using~\textbf{Algorithm~\ref{algoGP}}\;

$\mathbf{X}^{\star}\leftarrow\mathbf{X}_{n}$, $\boldsymbol{\theta}^{\star}\leftarrow\boldsymbol{\theta}_{n}$,
$\mathbf{\varsigma}^{\star}\leftarrow\mathbf{\varsigma}_{n}$\;

$\nu\leftarrow\nu+\omega f(\mathbf{X}^{\star},\boldsymbol{\theta}^{\star},C_{\mathrm{th}},\varsigma^{\star})$\;

$\omega\leftarrow\omega/\eta$\;

}
\end{algorithm}

\subsection{Complexity Analysis}

In this subsection, we present the complexity analysis of our proposed
PDDAGP method. In this context, we use $\mathcal{O}(\cdot)$ notation
to present the per-iteration complexity of~\textbf{Algorithm~\ref{algoPDDAGP}},
where we count the total number of required complex-valued multiplications.
It is important to note that the per-iteration complexity of~\textbf{Algorithm~\ref{algoPDDAGP}}
is dominated by that of the~\textbf{Algorithm~\ref{algoGP}}.

We first calculate the complexity of computing $\mathbf{X}_{n}$ (line~3)
in~\textbf{Algorithm~\ref{algoGP}}. For this purpose we need to
compute $\nabla_{\mathbf{X}}\hat{\mathsf{E}}_{\nu,\omega}(\mathbf{X}_{n-1},\boldsymbol{\theta}_{n-1},\varsigma_{n-1})$.whose
computational complexity is given by $\mathcal{O}(N_{\mathrm{B}}^{3}+N_{\mathrm{E}}^{3}+N_{\mathrm{B}}N_{\mathrm{A}}^{2}+N_{\mathrm{E}}N_{\mathrm{A}}^{2}+N_{\mathrm{B}}^{2}N_{\mathrm{A}}+N_{\mathrm{E}}^{2}N_{\mathrm{A}}+N_{\mathrm{S}}N_{\mathrm{A}}N_{\mathrm{B}}+N_{\mathrm{S}}N_{\mathrm{A}}N_{\mathrm{E}})$.
The complexity for obtaining an appropriate value of $\tau_{n}$ and
projecting $\hat{\mathbf{X}}_{n}$ onto $\mathcal{S}_{\mathbf{X}}$
results in an additional complexity of $\mathcal{O}(N_{\mathrm{A}}^{3})$.

Next, we calculate the complexity associated with the computation
of $\boldsymbol{\theta}_{n}$ (line~4) in~\textbf{Algorithm~\ref{algoGP}}.
Note that the complexity associated with the computation of an appropriate
$\mu_{n}$ and the projection of $\hat{\boldsymbol{\theta}}_{n}$
will be negligible as compared to that of computing $\nabla_{\boldsymbol{\theta}}\hat{\mathsf{E}}_{\nu,\omega}(\mathbf{X}_{n},\boldsymbol{\theta}_{n-1},\varsigma_{n-1})$,
and therefore, the complexity of computing $\boldsymbol{\theta}_{n}$
will be the same as that of $\nabla_{\boldsymbol{\theta}}\hat{\mathsf{E}}_{\nu,\omega}(\mathbf{X}_{n},\boldsymbol{\theta}_{n-1},\varsigma_{n-1})$,
given by $\mathcal{O}(N_{\mathrm{B}}N_{\mathrm{A}}^{2}+N_{\mathrm{E}}N_{\mathrm{A}}^{2}+N_{\mathrm{S}}N_{\mathrm{A}}^{2}+N_{\mathrm{A}}N_{\mathrm{B}}^{2}+N_{\mathrm{S}}N_{\mathrm{B}}^{2}+N_{\mathrm{S}}N_{\mathrm{E}}^{2}+N_{\mathrm{A}}N_{\mathrm{E}}^{2}+N_{\mathrm{S}}N_{\mathrm{A}}N_{\mathrm{B}}+N_{\mathrm{S}}N_{\mathrm{A}}N_{\mathrm{E}})$.

The complexity of computing $\varsigma_{n}$ (line~5) in~\textbf{Algorithm~\ref{algoGP}}
is negligible compared to that of $\mathbf{X}_{n}$ and $\boldsymbol{\theta}_{n}$.
Therefore, the per-iteration complexity of the Algorithm~\textbf{Algorithm~\ref{algoPDDAGP}}
is given by $\mathcal{O}(N_{\mathrm{A}}^{3}+N_{\mathrm{B}}^{3}+N_{\mathrm{E}}^{3}+N_{\mathrm{B}}N_{\mathrm{A}}^{2}+N_{\mathrm{E}}N_{\mathrm{A}}^{2}+N_{\mathrm{S}}N_{\mathrm{A}}^{2}+N_{\mathrm{A}}N_{\mathrm{B}}^{2}+N_{\mathrm{S}}N_{\mathrm{B}}^{2}+N_{\mathrm{A}}N_{\mathrm{E}}^{2}+N_{\mathrm{S}}N_{\mathrm{A}}N_{\mathrm{B}}+N_{\mathrm{S}}N_{\mathrm{A}}N_{\mathrm{E}})$.
We note that in a practical IRS-MIMOME system we should have $N_{\mathrm{S}}\gg\max\{N_{\mathrm{A}},N_{\mathrm{B}},N_{\mathrm{E}}\}$.
Therefore, the complexity of the proposed algorithm can be approximated
by $\mathcal{O}(N_{\mathrm{S}}N_{\mathrm{A}}(N_{\mathrm{B}}+N_{\mathrm{E}}))$,
meaning that the computational complexity of the proposed PDDAGP algorithm
increases linearly with $N_{\mathrm{S}}$. We remark that for the
IRS-assisted MISO WTC case, the complexity of the proposed algorithm
is reduced to to $\mathcal{O}(N_{\mathrm{S}}N_{\mathrm{A}})$, which
is notably lower than that of the SDP-based algorithm proposed in~\cite{Secure_EE_Jamming}
and the former is more suitable for practical implementation.

\begin{figure*}[t]
\noindent\begin{minipage}[t]{0.30\linewidth}%
\begin{center}
\includegraphics[width=0.9\columnwidth]{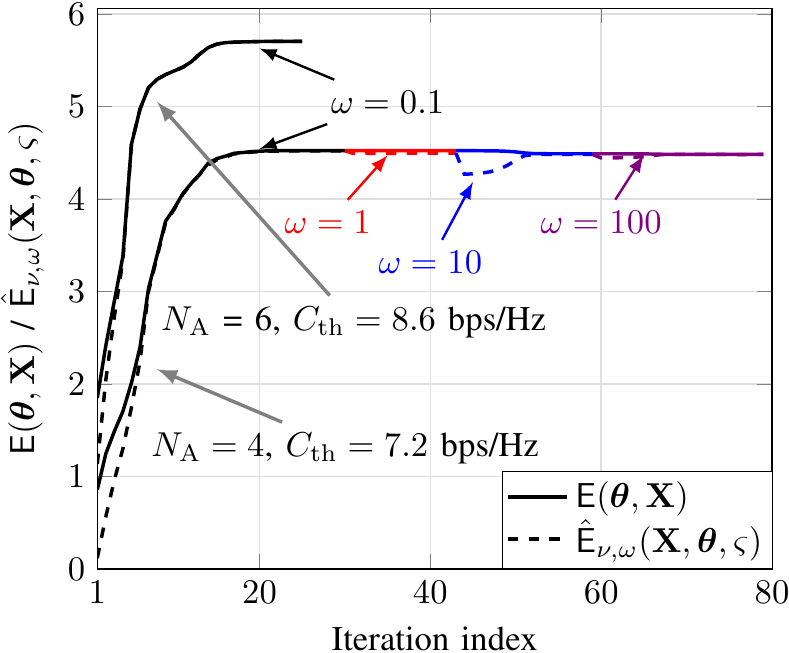}
\par\end{center}
\caption{Convergence results of the PDDAGP method for $(N_{\mathrm{B}},N_{\mathrm{E}},N_{\mathrm{S}})=(4,4,100)$.}

\label{fig:convergence}%
\end{minipage}\hfill{}%
\noindent\begin{minipage}[t]{0.30\linewidth}%
\begin{center}
\includegraphics[width=0.9\columnwidth]{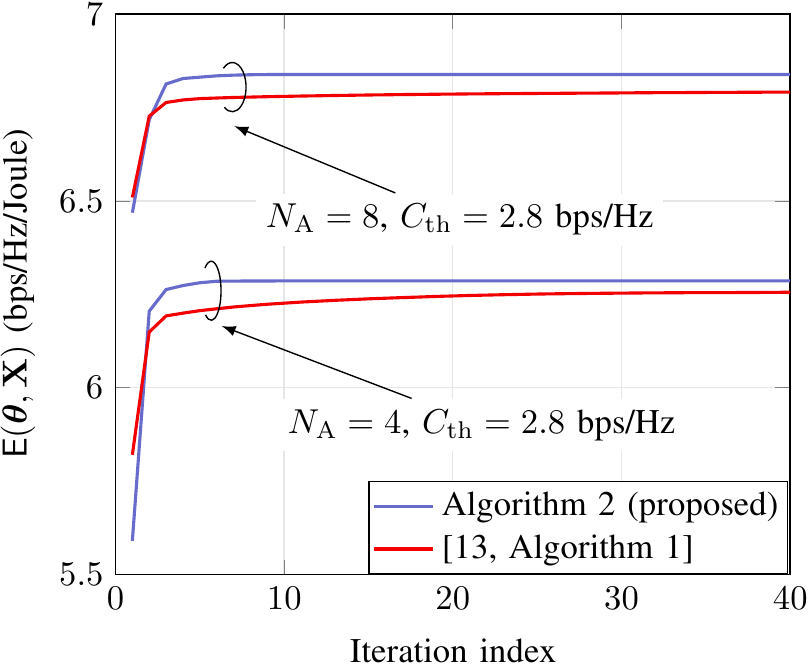}
\par\end{center}
\caption{Convergence comparison for $(N_{\mathrm{B}},N_{\mathrm{E}},N_{\mathrm{S}})=(1,1,64)$.}

\label{fig:MISOConvergenceComparison}%
\end{minipage}\hfill{}%
\noindent\begin{minipage}[t]{0.30\linewidth}%
\begin{center}
\includegraphics[width=0.9\columnwidth]{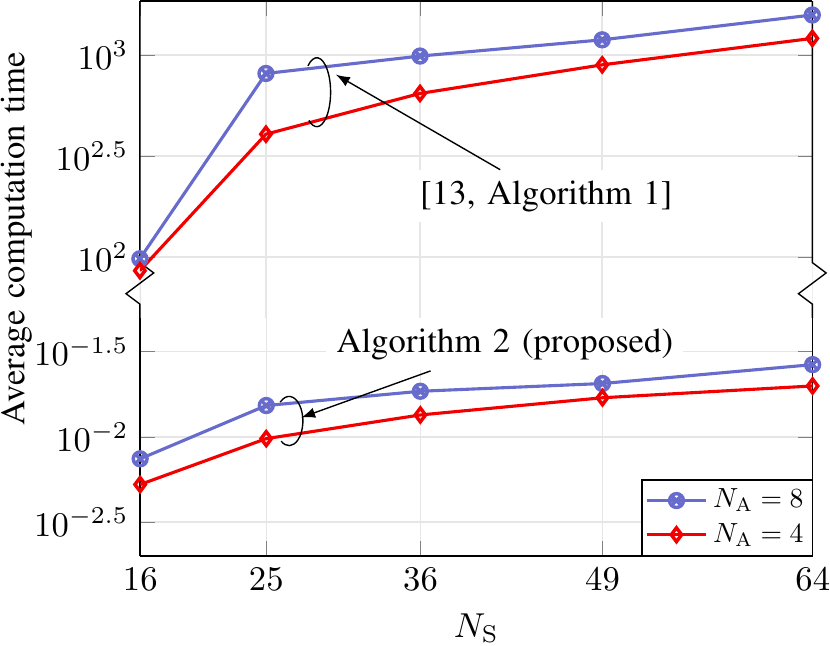}
\par\end{center}
\caption{Average computation time for various algorithms for $(N_{\mathrm{B}},N_{\mathrm{E}})=(1,1)$.}

\label{fig:RunTimeSDRvsPDD}%
\end{minipage}
\end{figure*}

\section{Numerical Analysis}

In this section, we first describe the simulation parameters considered
in this paper. The center of the Alice's, Bob's and Eve's uniform
linear array is assumed to be located at $(0,l_{\mathrm{A}},h_{\mathrm{A}})$,
$(d_{\mathrm{B}},l_{\mathrm{B}},h_{\mathrm{B}})$, and $(d_{\mathrm{E}},l_{\mathrm{E}},h_{\mathrm{E}})$,
respectively. The center of the IRS is denoted by $(d_{\mathrm{S}},0,h_{\mathrm{S}})$.
The inter-antenna separation at Alice, Bob, and Eve, and the distance
between any two of the neighboring reflecting elements at IRS is assumed
to be equal to $\lambda/2$; here $\lambda$ denotes the wavelength
of the carrier waves.

The Alice-Bob and Alice-Eve channels are defined as $\mathbf{H}_{\mathrm{AJ}}=\sqrt{(\kappa+1)^{-1}}\boldsymbol{\Xi}_{\mathrm{AJ}}\odot\bigl(\sqrt{\kappa}\mathbf{H}_{\mathrm{AJ},\mathrm{LoS}}+\mathbf{H}_{\mathrm{AJ,}\textrm{NLoS}}\bigr)$,
$\mathrm{J}\in\{\mathrm{B,E}\}$, where the elements in $\mathbf{H}_{\mathrm{AJ},\mathrm{LoS}}$
given by $e^{-j2\pi l_{p,q}/\lambda}$ corresponds to the line-of-sight
(LoS) component, $\mathbf{H}_{\mathrm{AJ},\mathrm{NLoS}}\sim\mathcal{CN}\left(\mathbf{0},\mathbf{I}\right)$
represents the non-line-of-sight (NLoS) component, $l_{p,q}$ denotes
the distance between the $p$-th antenna at the transmitter (Alice)
and the $q$-th antenna at the receiver (Bob/Eve), and $\kappa$ is
the Rician factor. Moreover, the elements in $\boldsymbol{\Xi}_{\mathrm{AJ}}\in\mathbb{R}^{N_{\mathrm{J}}\times N_{\mathrm{A}}}$
correspond to the free-space path loss (FSPL) coefficients and are
given by $((4\pi/\lambda)^{2}l_{p,q}^{3})^{-1/2}$. Similarly, the
IRS-Bob and IRS-Eve links are modeled as $\mathbf{H}_{\mathrm{SJ}}=\sqrt{(\kappa+1)^{-1}}\boldsymbol{\Xi}_{\mathrm{SJ}}\odot(\sqrt{\kappa}\mathbf{H}_{\mathrm{SJ},\mathrm{LoS}}+\mathbf{H}_{\mathrm{SJ},\mathrm{NLoS}})$.
The elements in $\mathbf{H}_{\mathrm{SJ},\mathrm{LoS}}$ are given
by $e^{-j2\pi l_{k,q}/\lambda}$, $\mathbf{H}_{\mathrm{SJ,}\mathrm{NLoS}}\sim\mathcal{CN}\left(\mathbf{0},\mathbf{I}\right)$,
the elements in $\boldsymbol{\Xi}_{\mathrm{SJ}}$ are given by $(\Upsilon_{\mathrm{J}}l_{\mathrm{J}}(4\pi/\lambda)^{2}l_{k,q}^{3})^{-1/2}$,
$l_{k,q}$ is the distance between the $k$-th reflecting element
at the IRS and the $q$-th antenna at Bob/Eve and $\Upsilon_{\mathrm{J}}=2$
is the antenna gain at Bob/Eve. The Alice-IRS channel is modeled as
$\mathbf{H}_{\mathrm{AS}}=\sqrt{(\kappa+1)^{-1}}\boldsymbol{\Xi}_{\mathrm{AS}}\odot(\sqrt{\kappa}\mathbf{H}_{\mathrm{AS},\mathrm{LoS}}+\mathbf{H}_{\mathrm{AS},\mathrm{NLoS}})$,
where the elements in $\boldsymbol{\Xi}_{\mathrm{AS}}$ given by $(\Upsilon_{\mathrm{A}}l_{\mathrm{A}}(4/\lambda)^{2}l_{p,k}^{3})^{-1/2}$
represent the FSPL coefficients, the elements in $\mathbf{H}_{\mathrm{AS},\mathrm{LoS}}$
are given by $e^{-j2\pi l_{p,k}/\lambda}$, $\mathbf{H}_{\mathrm{AS,NLoS}}\sim\mathcal{CN} \left(\mathbf{0},\mathbf{I}\right)$,
$l_{p,k}$ is the distance between the $p$-th antenna at Alice and
the $k$-th reflecting element at the IRS, and $\Upsilon_{\mathrm{A}}=2$
is the antenna gain at Alice.

\begin{figure*}[t]
\noindent\begin{minipage}[t]{0.29\linewidth}%
\begin{center}
\includegraphics[width=0.93\columnwidth]{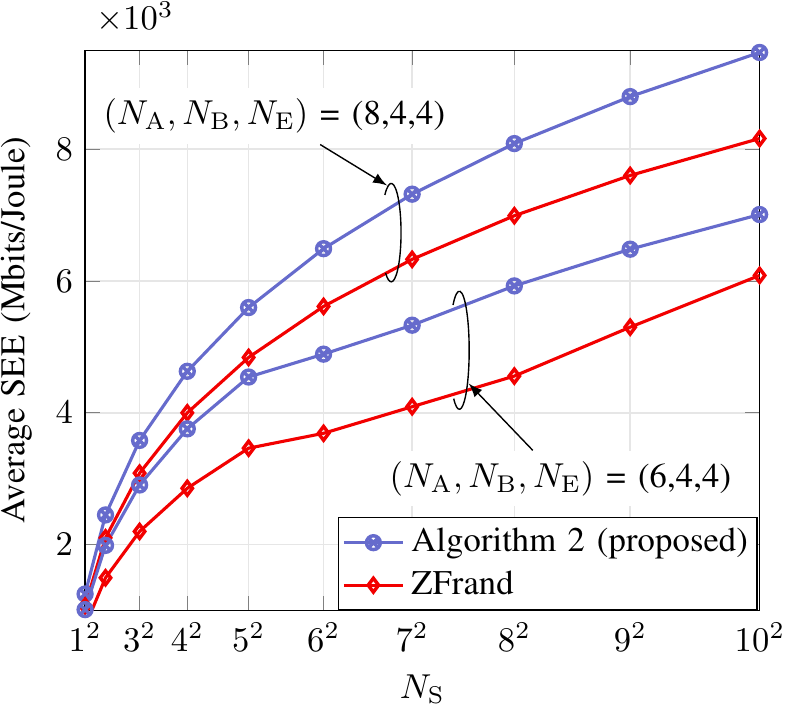}
\par\end{center}
\caption{The average SEE versus $N_{\mathrm{S}}$ for $C_{\mathrm{th}}=1.4$
bps/Hz.}

\label{fig:PDDvsZF}%
\end{minipage}\hfill{}%
\noindent\begin{minipage}[t]{0.29\linewidth}%
\begin{center}
\includegraphics[width=0.93\columnwidth]{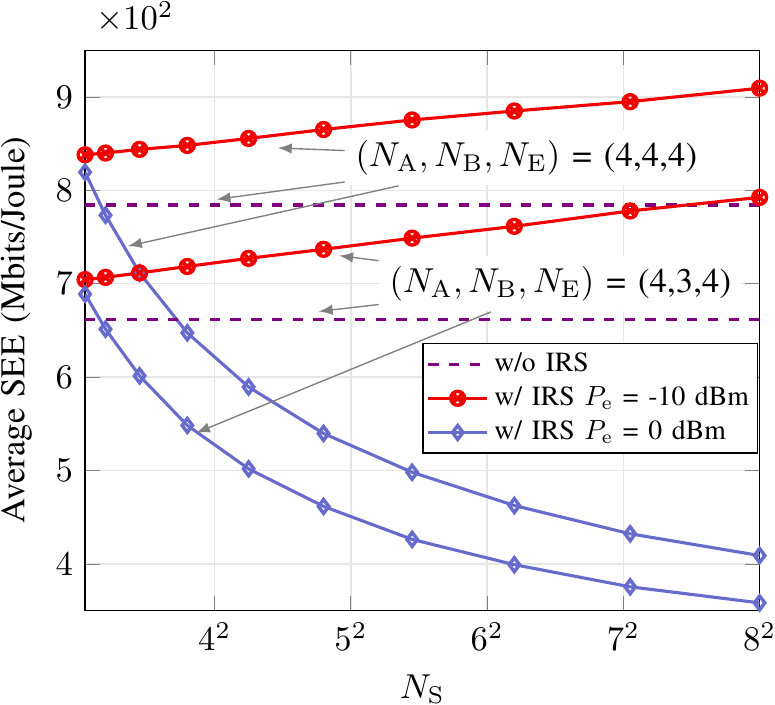}
\par\end{center}
\caption{The average SEE versus $N_{\mathrm{S}}$ for $C_{\mathrm{th}}=0.14$
bps/Hz.}

\label{fig:RISImpact}%
\end{minipage}\hfill{}%
\noindent\begin{minipage}[t]{0.29\linewidth}%
\begin{center}
\includegraphics[width=0.93\columnwidth]{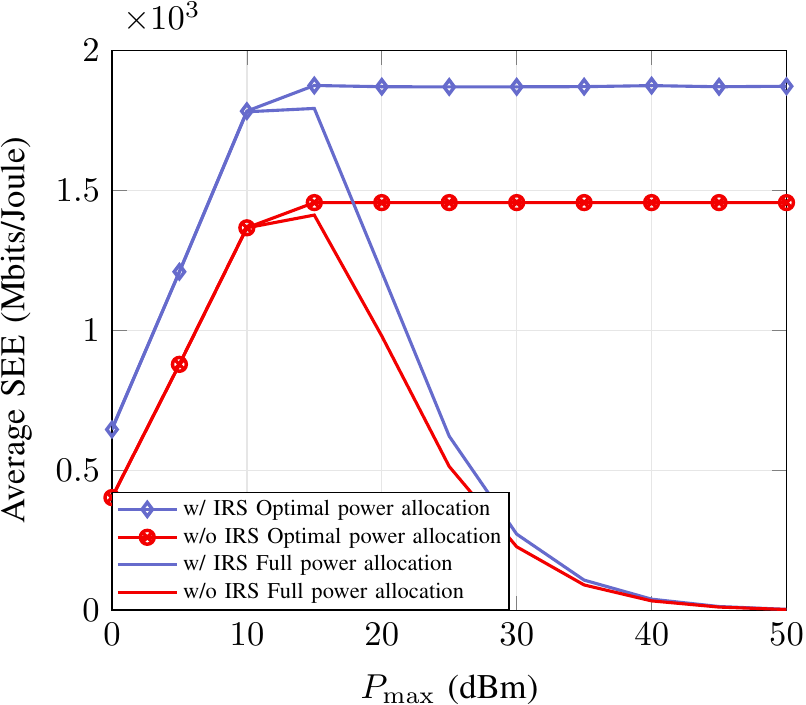}
\par\end{center}
\caption{The average SEE versus $P_{\max}$ for $(N_{\mathrm{A}},N_{\mathrm{B}},N_{\mathrm{E}},N_{\mathrm{S}})=(8,6,6,100)$
for $C_{\mathrm{th}}=0.14$ bps/Hz.}

\label{fig:TxPowImp}%
\end{minipage}
\end{figure*}

Unless stated otherwise, in this section we consider $\kappa=1$,
$l_{\mathrm{A}}=20$~m, $h_{\mathrm{A}}=10$~m, $d_{\mathrm{B}}=350$~m,
$l_{\mathrm{B}}=10$~m, $h_{\mathrm{B}}=2$~m, $d_{\mathrm{E}}=352$~m,
$l_{\mathrm{E}}=15$~m, $h_{\mathrm{E}}=2$~m, $d_{\mathrm{S}}=30$~m,
$h_{\mathrm{S}}=5$~m, $P_{\max}=40$~dBm, $\mathscr{N}_{0}=$ $-174$~dBm/Hz,
$\alpha=0.833$, $B=20$~MHz, $\lambda=0.15\ \mathrm{m}$, $P_{\mathrm{A}}=P_{\mathrm{B}}=10$~dBm
and $P_{\mathrm{e}}=0.01$~dBm. We performed the numerical experiments
using the MATLAB (R2022a) on a 64-bit Windows machine with 16 GB RAM
and an Intel Core i7 3.20 GHz processor. Note that in Fig.~\ref{fig:RunTimeSDRvsPDD}-\ref{fig:TxPowImp},
the average rum time/average SEE is computed over 1000 independent
channel realizations and the Fig.~\ref{fig:convergence},\ref{fig:MISOConvergenceComparison}
are constructed based on single channel realization.

We now present the numerical results to evaluate the performance of
the proposed method for the IRS-MIMOME system under consideration.
In Fig. \ref{fig:convergence}, we show the convergence results of~\textbf{Algorithm~\ref{algoPDDAGP}}.
It is evident from Fig.~\ref{fig:convergence} that the augmented
objective function $\hat{\mathsf{E}}_{\nu,\omega}(\mathbf{X},\boldsymbol{\theta},\varsigma)$
and the original objective function $\mathsf{E}(\mathbf{X},\boldsymbol{\theta})$
converge to the same value, indicating that the regularized term becomes
zeros at convergence, which is in-line with the theory of penalty-based
optimization schemes. Note that for fixed $\nu$ and $\omega$, we
consider $\hat{\mathsf{E}}_{\nu,\omega}(\mathbf{X},\boldsymbol{\theta},\varsigma)$
to be converged when the relative difference in $\hat{\mathsf{E}}_{\nu,\omega}(\mathbf{X},\boldsymbol{\theta},\varsigma)$,
i.e., $[\hat{\mathsf{E}}_{\nu,\omega}(\mathbf{X}_{n},\boldsymbol{\theta}_{n},\varsigma_{n})-\hat{\mathsf{E}}_{\nu,\omega}(\mathbf{X}_{n-1},\boldsymbol{\theta}_{n-1},\varsigma_{n-1})]/\hat{\mathsf{E}}_{\nu,\omega}(\mathbf{X}_{n-1},\boldsymbol{\theta}_{n-1},\varsigma_{n-1})$
becomes less than or equal to the tolerance value $(\epsilon=10^{-4})$.
After this, we update the Lagrangian multiplier $\nu$ and the penalty
parameter $\omega$. Finally the algorithm is considered convergent
when $|\mathsf{E}(\mathbf{X}_{n},\boldsymbol{\theta}_{n})-\hat{\mathsf{E}}_{\nu,\omega}(\mathbf{X}_{n},\boldsymbol{\theta}_{n},\varsigma_{n})|/\hat{\mathsf{E}}_{\nu,\omega}(\mathbf{X}_{n},\boldsymbol{\theta}_{n},\varsigma_{n})\leq\epsilon$.
We remark that~\textbf{Algorithm~\ref{algoPDDAGP}} aims to maximize
the augmented objective function $\hat{\mathsf{E}}_{\nu,\omega}(\mathbf{X},\boldsymbol{\theta},\varsigma)$,
and thus for fixed values of $\nu$ and $\omega$, $\hat{\mathsf{E}}_{\nu,\omega}(\mathbf{X},\boldsymbol{\theta},\varsigma)$
increases monotonically.

In Fig.~\ref{fig:MISOConvergenceComparison} we compare the performance
our proposed algorithm with that of~\cite[Algorithm 1]{Secure_EE_Jamming}.
It is important to note that since~\cite[Algorithm 1]{Secure_EE_Jamming}
is applicable only for IRS-MISOSE system, in~Fig.~\ref{fig:MISOConvergenceComparison}
we consider $N_{\mathrm{B}}=N_{\mathrm{E}}=1$. It is evident from
the figure that both the proposed PDDAGP and~\cite[Algorithm 1]{Secure_EE_Jamming}
takes almost the same number of iterations to achieve the convergence,
however, the proposed PDDAGP method attains a higher SEE. The main
reason for the inferior performance of the method proposed in~\cite[Algorithm 1]{Secure_EE_Jamming},
which is based on AO, is its way in handling the coupling between
the optimization variables in the constraint. It is well known that
AO-based algorithms can be easily trapped in ineffective solution
due to poor initialization. The involved Gaussian randomization is
another reason for the inferior performance of~\cite[Algorithm 1]{Secure_EE_Jamming}.
Note that in the case of PDDAGP algorithm, the coupling in the constraint
is brought to the objective in the form of a penalized term. This
move makes optimization variables decoupled, which results in a superior
performance as demonstrated. 

We now numerically shown that the proposed algorithm also outperforms
~\cite[Algorithm 1]{Secure_EE_Jamming} in terms of the required
runtime, which validates the complexity analysis presented in preceding
section. To this end, in Fig.~\ref{fig:RunTimeSDRvsPDD} we compare
the average runtime of \textbf{Algorithm~\ref{algoPDDAGP}} and~\cite[Algorithm 1]{Secure_EE_Jamming}
w.r.t. $N_{\mathrm{S}}$. It is clearly evident from the figure that
the proposed PDDAGP algorithm takes significantly less time to converge
on average compared to that for~\cite[Algorithm 1]{Secure_EE_Jamming}.
This is because the complexity of our proposed algorithm grows only
linearly w.r.t. $N_{\mathrm{S}}$, where the complexity of SDP-assisted
Dinkelbach's method in~\cite[Algorithm 1]{Secure_EE_Jamming} is
excessively high.

In Fig.~\ref{fig:PDDvsZF}, we compare the average SEE performance
versus $N_{\mathrm{S}}$ for the proposed PDDAGP algorithm with that
of a baseline scheme which we refer to as ZFrand. In this ZFrand scheme,
for a fixed $\boldsymbol{\theta}$ we find $\mathbf{X}\in\mathcal{S}_{\mathbf{X}}$
that maximizes~(\ref{eq:mainOptProb}) while satisfying $\mathbf{H}_{\mathrm{E}}\mathbf{X}\mathbf{H}_{\mathrm{E}}\herm=\boldsymbol{0}$
(which is essentially the zero-forcing precoder) and~(\ref{eq:SecThresConstraint}).
Due to the ZF constraint, problem (\ref{eq:mainOptProb}) can be reformulated
as a convex program and thus $\mathbf{X}$ can be found exactly. Then
for a given $\mathbf{X}$, to update $\boldsymbol{\theta}$, we follow
a procedure similar to Gaussian randomization, i.e., randomly many
$\boldsymbol{\theta}$s and select the one that gives the best objective
. It is noticeable that our proposed PDDAGP method outperforms the
ZFrand scheme, which is expected because of the suboptimality of ZF
beamformer and as well as that of selecting the random $\boldsymbol{\theta}$.

In Fig.~\ref{fig:RISImpact}, we show the impact of the number of
reflecting elements at the IRS, i.e., $N_{\mathrm{S}}$. Note that
increasing $N_{\mathrm{S}}$ increases both $C(\mathbf{X},\boldsymbol{\theta})$
and $P_{\mathrm{total}}$, resulting in a non-trivial trade-off for
the SEE. In the figure, we consider different values for the circuit
power consumption per IRS element, i.e., $P_{\mathrm{e}}$. It is
observed from Fig.~\ref{fig:RISImpact} that for a small value of
$P_{\mathrm{e}}$, the IRS-MIMOME system can achieve a significantly
higher SEE compared to its non-IRS counterpart. On the other hand,
a large value of $P_{\mathrm{e}}$ may result in a performance degradation
in terms of average SEE in an IRS-MIMOME system compared to the non-IRS
counterpart. Therefore, it can be concluded that for a given $P_{\mathrm{e}}$,
there exists an optimal value of $N_{\mathrm{S}}$ to maximize the
benefit of using an IRS in terms of average SEE.

In Fig.~\ref{fig:TxPowImp}, we compare the average SEE of the system
under the consideration with its non-IRS counterpart, for two different
scenarios: (i) optimal power allocation and optimal IRS phase-shifts
(i.e.,~(\ref{eq:mainOptProb})), and (ii) full power allocation and
optimal IRS phase-shifts (i.e.,~(\ref{eq:mainOptProb}) with~(\ref{eq:TPC})
modified to $\tr(\mathbf{X})=P_{\max}$). We note that $C(\mathbf{X},\boldsymbol{\theta})$
is the logarithmic function while $P_{\mathrm{total}}$ is a linear
function w.r.t. to $\mathbf{X}$. So, in the case of full power allocation,
for small values of $P_{\max}$, the average SEE first increases with
increasing values of $P_{\max}$ because the rate of increase of $C(\mathbf{X},\boldsymbol{\theta})$
is larger than that of the $P_{\mathrm{total}}$. However, for large
values of $P_{\max}$, $P_{\mathrm{total}}$ increases at a much faster
rate compared to $C(\mathbf{X},\boldsymbol{\theta})$, resulting in
a loss in SEE. On the other hand, for the case of optimal power allocation,
the system uses full power ($P_{\max}$) for small values of $P_{\max}$,
however, it uses a smaller transmit power than the maximum available
$P_{\max}$ for large value of $P_{\max}$. Specifically, for the
case of optimal power allocation in the large $P_{\max}$ regime,
$\tr(\mathbf{X})$ becomes strictly less than $P_{\max}$, irrespective
of the value of $P_{\max}$ which results in a constant value of both
$C(\mathbf{X},\boldsymbol{\theta})$ and $P_{\mathrm{total}}$. This
in turn results in a saturated value of SEE, i.e., $\mathsf{E}(\mathbf{X},\boldsymbol{\theta})$
for increasing value of $P_{\max}$.

\section{Conclusion}

In this paper, we considered the problem of the SEE maximization in
an IRS-MIMOME system, subject to a transmit power constraint (at the
transmitter), a target secrecy rate constraint (at the legitimate
receiver), and unit-modulus constraints at the IRS. We proposed a
PDDAGP method to find a stationary solution to the challenging non-convex
optimization problem of the SEE maximization. Extensive numerical
experiments were performed to evaluate the performance of the proposed
algorithm, and the superiority of the PDDAGP method was also established
over that of an SDP-assisted Dinkelbach's method for the special case
of IRS-MISOSE system. \textit{\emph{We also showed that the per-iteration
complexity of the proposed PDDAGP method grows as linearly w.r.t.
the number of reflecting elements at the IRS,}} that is appealing
for practical implementation.

\bibliographystyle{IEEEtran}
\bibliography{paper}

\end{document}